\documentclass[twocolumn,aps,superscriptaddress,showpacs,nofootinbib,floatfix]{revtex4-1}
\usepackage{epsfig,bm,feynmf}
\usepackage{graphics}
\usepackage{amsmath}
\usepackage[eps]{pstricks}
\usepackage[normalem]{ulem}  

\renewcommand\sout{\bgroup \color{red} \ULdepth=-.5ex \ULset}

\newcommand{\als}{\alpha_{s}}

\newcommand{\gev}{\text{GeV}}

\begin{document}


\title{Vector meson mass in the chiral symmetry restored vacuum}


\author{Jisu Kim}%
\email{fermion0514@yonsei.ac.kr}
\affiliation{Department of Physics and Institute of Physics and Applied Physics, Yonsei University, Seoul 03722, Korea}

\author{Su Houng Lee}%
\email{suhoung@yonsei.ac.kr}
\affiliation{Department of Physics and Institute of Physics and Applied Physics, Yonsei University, Seoul 03722, Korea}


\begin{abstract}
We calculate the mass of the vector meson in the chiral symmetry restored vacuum.  This is accomplished by separating the four quark operators appearing in the vector and axial vector meson sum rules into chiral symmetric and symmetry breaking parts depending on the contribution of the fermion zero modes.  We then identify each part from the fit to the vector and axial vector meson masses.  By taking the chiral symmetry breaking part to be zero while keeping the symmetric operators to the vacuum value, we find that the chiral symmetric part of the vector and axial vector meson mass to be between 550 and  600 MeV.  This demonstrates that chiral symmetry breaking, while responsible for the mass difference between chiral partner, accounts only for a small fraction of the symmetric part of the mass.  
\end{abstract}


\maketitle

\section{Introduction}

Explaining the mass of a hadron  from the  underlying QCD dynamics is one of the fundamental problems in QCD\cite{Wilczek:1999be,Wilczek:2012sb}.  
The phenomenological success of constituent quark model or effective chiral models   lead us to believe that spontaneous chiral symmetry breaking\cite{Nambu:1961tp,Nambu:1961fr} is partly responsible for generating hadron masses \cite{Hatsuda:1985eb,Brown:1991kk,Hatsuda:1991ez,Leupold:2009kz}.

As a means to understand the origin of hadron mass, experiments have been performed worldwide to observe mass shift of hadrons at finite temperature or density \cite{Hayano:2008vn,Metag:2017yuh,Ichikawa:2018woh}.  
This is so because the initial condition at heavy ion collision is expected to probe the quark-gluon plasma phase where chiral symmetry is expected to be restored.  Furthermore, nuclear target experiments can probe the nuclear matter environment, where the order parameter is known to decease by almost 30$\%$.  Hence, measuring the property change of hadron related to chiral symmetry restoration effects is one of the future aims at  Hadron physics program at J-PARC\cite{Ohnishi:2019cif} and heavy ion programs world wide\cite{Salabura:2020tou}. 

The mass difference between the chiral partners are known to be due to the chiral symmetry breaking\cite{Weinberg:1967kj}.  
Hence, if chiral symmetry is restored, the mass difference between chiral partners should vanish.  For example, the $K^*$ and $K_1$ meson masses will  become degenerate if chiral symmetry gets restored, which could be observed due to their small vacuum width\cite{Song:2018plu,Lee:2019tvt}. 
However, with confinement phenomena and scale breaking in QCD\cite{Collins:1976yq}, how much of the total hadron mass comes from chiral symmetry breaking is still far from being understood.

In the original in-medium QCD sum rules for the light vector mesons\cite{Hatsuda:1991ez},
the changes of the masses were dominantly due to the change of the  four-quark condensate, whose medium expectation value was extracted under the so called vacuum saturation hypothesis so that it is proportional to the square of the quark condensate\cite{Shifman:1978bx}.  Hence, in this approximation, the four quark condensate vanishes if the quark condensate vanishes leading to a small vector meson mass sustained only by the small contribution from the gluon condensate.  
However, it is not clear if all the four quark operators vanish if chiral symmetry is restored as only the difference between the four quark operators appearing in the vector and axial vector meson sum rule is an order parameter of chiral symmetry breaking.  

Models for the vector mesons based on chiral symmetry satisfy the Weinberg type relation for the vector and axial vector meson mass.  But the common mass has to be assumed and it is not clear how that part changes when chiral symmetry is restored.  The question of whether chiral symmetry breaking is the origin of the hadron mass can be answered by isolating the chiral symmetry breaking effects in the vacuum.  In the past, a similar question has been addressed using the lattice gauge theory, where the hadronic correlation functions have been studied after the  lattice cooling process\cite{Chu:1994vi} that eleminated the short distance Coulomb and confinement physics.  

In this work, we calculate the mass of the vector and axial vector meson mass 
in the chiral symmetry restored vacuum.  This is accomplished by identifying the chiral symmetric and breaking part of the four quark operators through their dependencies on the fermion zero mode\cite{BC} and then estimating their magnitudes by fitting to the vector and axial vector meson mass via QCD sum rule method. 
Applying the QCD sum rule approach when the chiral symmetry breaking operator is taken to be zero while keeping  the chiral symmetric four quark operator to its vacuum value, we find that that the vector meson mass  becomes  about 550 to  600 MeV. 
Our result suggest that chiral symmetry breaking is responsible for the mass difference between chiral partners but has only a small contribution to the common vector and axial vector meson mass.  

{\it Operator product expansion (OPE) for vector and axial vector current: }
The commonly known chiral order parameter can be rewritten in several forms.
\begin{eqnarray}
\langle \bar{q} q \rangle &  = &  \lim_{x \rightarrow 0} 
- \frac{1}{2} \langle {\rm Tr} [ S(0,x)- i\gamma_5  S(0,x) i \gamma_5 ] ] \rangle  \nonumber \\ 
&  = & -\pi \langle \rho(\lambda=0) \rangle, 
\label{BC-0}
\end{eqnarray}
where the second line shows the density of zero eigenvalue in the Euclidean formalism known as the Banks-Casher formula\cite{BC}.  
The formula is useful as it identifies the origin of chiral symmetry breaking, and further can be used to isolate the chiral symmetry breaking part of any quark operator: a set of order parameters  can be obtained by looking at the difference in the four quark operators\cite{Cohen:1996ng,Lee:2019tvt}.

In the two point correlation functions of the vector current $J^\rho_\mu =\bar{q} \tau^3 \gamma_\mu q$ and the axial vector current $J^{a_1}_\mu =\bar{q} \tau^3 \gamma_\mu \gamma^5 q$ current, the relevant matrix elements of dimension 6 operators contributing as ${\cal M}/Q^6$ to the transverse part of the correlation function $\Pi=\Pi_\mu^\mu/(-3q^2)$, in the SU(2) flavor limit, are respectively given for the $\rho$ and $a_1$ channel as\cite{Shifman:1978bx}
\begin{eqnarray}
{\cal M}_\rho  & = & - 2 \pi \alpha_s 
\langle ( \bar{q} \gamma_\mu \gamma^5 \lambda^a \tau^3 q )^2 \rangle 
\nonumber \\ 
&&
-\frac{4 \pi \alpha_s}{9} 
\langle (\sum_{ud} \bar{q} \gamma_\mu \lambda^a  q )(\sum_{uds} \bar{q} \gamma_\mu \lambda^a  q ) \rangle,
\label{rho} \\
{\cal M}_{a_1} & =  & -
2 \pi \alpha_s
\langle ( \bar{q} \gamma_\mu  \lambda^a \tau^3 q )^2 \rangle
\nonumber \\
&&
- \frac{4 \pi \alpha_s}{9} 
\langle (\sum_{ud} \bar{q} \gamma_\mu \lambda^a  q )(\sum_{uds} \bar{q} \gamma_\mu \lambda^a  q ) \rangle .
\label{a1}
\end{eqnarray}

For the first operators appearing in both channels, one can define the following operators
\begin{eqnarray}
\langle ( \bar{q} \gamma_\mu \gamma^5 \lambda^a \tau^3 q )^2 \rangle_{S,B}  =  
\frac{1}{2} 
\langle ( \bar{q} \gamma_\mu \gamma^5 \lambda^a \tau^3 q )^2 
\pm ( \bar{q} \gamma_\mu  \lambda^a \tau^3 q )^2 \rangle, 
\nonumber  \\
\langle ( \bar{q} \gamma_\mu  \lambda^a \tau^3 q )^2 \rangle_{S,B}  =  
\frac{1}{2} 
\langle ( \bar{q} \gamma_\mu  \lambda^a \tau^3 q )^2 
\pm ( \bar{q} \gamma_\mu \gamma^5 \lambda^a \tau^3 q )^2 \rangle, 
\nonumber  \\ \label{decomposition}
\end{eqnarray}
where the subscript $S,B$ refers to chiral symmetric and breaking operators.  The fermion zero modes only contribute to the chiral symmetry breaking operator, which therefore constitute a chiral order parameter.
  
As for the last operators appearing in Eq.~(\ref{rho}) and in Eq.~(\ref{a1}), one notes that they are identical.  Furthermore, performing chiral symmetry transformation would leave the operator invariant.  However, spontaneous chiral symmetry will introduce non-vanishing contribution to the expectation value of this operator.  This is so because the zero modes contributes to this operator.
Keeping this in mind, we rewrite the non-strangeness part as
\begin{eqnarray}
\langle ( \bar{q} \gamma_\mu \lambda^a  q )^2 \rangle  & = &
\langle ( \bar{q} \gamma_\mu \lambda^a  q )^2 \rangle_d
+  \langle ( \bar{q} \gamma_\mu \lambda^a \tau^3 q )^2 \rangle,
\label{singlet-four}
\end{eqnarray}
where $
\langle ( \bar{q} \gamma_\mu \lambda^a  q )^2 \rangle_d = 
\langle ( \bar{q} \gamma_\mu \lambda^a  q )^2 \rangle- \langle ( \bar{q} \gamma_\mu \lambda^a \tau^3 q )^2 \rangle $ represents the disconnected contribution with no contribution from the fermion zero mode.
Here, $q$ runs over the $(u,d)$ quarks.     Eq.~(\ref{decomposition}) can be used in the last term of  Eq.~(\ref{singlet-four}) to extract the chiral symmetry breaking part.

Hence, we can rewrite the dimension 6 contributions as 
\begin{eqnarray}
{\cal M}_\rho & =  & 
\frac{14  }{9} B + S, 
 \nonumber \\
{\cal M}_{a_1} & =  & 
-\frac{22  }{9} B+   S, 
\end{eqnarray}
where
\begin{eqnarray}
B &  = &  -
\pi \alpha_s  \langle ( \bar{q} \gamma_\mu \gamma^5 \lambda^a \tau^3 q )^2 \rangle_B  \nonumber \\ 
S & = & -
\frac{22 \pi \alpha_s}{9} \langle ( \bar{q} \gamma_\mu \gamma^5 \lambda^a \tau^3 q )^2 \rangle_S   \nonumber \\
&& - \frac{4 \pi \alpha_s}{9} \bigg(
 \langle ( \bar{q} \gamma_\mu  \lambda^a  q )^2 \rangle_d  + \langle
 ( \bar{q} \gamma_\mu  \lambda^a  q )( \bar{s} \gamma_\mu  \lambda^a  s )
 \rangle_S  \bigg). 
\label{BS}
\end{eqnarray}
The ferimon zero mode only contributes in $B$ making it a chiral symmetry breaking operator.  The operators in  $S$ are chiral symmetric operators.  
One notes that the chiral symmetric operators are identical in the $\rho,a_1$ sum rule, while the chiral symmetry breaking operator contribute with different sign and coefficients.  This is expected as the chiral symmetry breaking operator $B$ is responsible for the $\rho-a_1$ difference, while the symmetric operators $S$ contribute universally to both sum rules. We will now use the QCD sum rule for $\rho$ and $a_1$ to determine the magnitude of both operators $B$ and $S$. 

{\it Sum rule analysis} 
One starts with the Borel transformed dispersion relation for the invariant part of the correlation function 
\begin{eqnarray}
\widehat{\Pi}(M^2) = \int_{0}^{\infty} ds e^{-s/M^2} \rho(s),
\label{sum_rules}
\end{eqnarray}
where $\widehat{\Pi}(M^2)$ represents the Borel transformed OPE  of the correlator $\Pi$ and $M$ stands for the Borel mass.  $\rho(s)$ is the spectral function for either the $\rho$ or $a_1$ meson. 
We will take the spectral function to have the following pole and continuum contribution
\begin{eqnarray}
\rho^{\mathrm{pole}}(s) &=& \frac{1}{\pi}
\frac{f \Gamma \sqrt{s}}{(s - m^2)^2 + s\Gamma^2},  \label{pheno_side_pole}
\\
\rho^{\mathrm{cont}}(s) &=& \frac{1}{\pi} \theta(s - s_0) \mathrm{Im} \tilde{\Pi}^{\mathrm{pert}}(s).
\label{pheno_side_2}
\end{eqnarray}
We substitute the spectral density into Eq.~(\ref{sum_rules}) and take the OPE up to dimension 6 operators.  We then take the ratio of the pole contribution to its derivative with respect to the inverse Borel mass to eliminate the dependence on $f$. For the $a_1$ meson sum rule, we take further derivative to eliminate the pion contribution\cite{Hatsuda:1992bv}. We can then obtain sum rules for the four quark condensate for the $\rho$ and $a_1$ meson. Normalized to the vacuum saturation hypothesis, we find
\begin{widetext}
\begin{eqnarray}
\kappa_\rho & = &  \bigg[ I^{(1)} \bigg( (1+\frac{\alpha_s}{\pi})E_0+ \frac{a+b}{M^4} \bigg) -I^{(0)}\bigg( (1+\frac{\alpha_s}{\pi})E_1 M^2- \frac{a+b}{M^2} \bigg) \bigg] \times \bigg[ 2I^{(0)}\frac{d_\rho}{M^4} + I^{(1)}\frac{d_\rho}{M^6} \bigg]^{-1}  
\nonumber \\
\kappa_{a_1} & = & \bigg[ -2 I^{(1)} \bigg( (1+\frac{\alpha_s}{\pi})E_2M^2 \bigg) + I^{(2)}\bigg( (1+\frac{\alpha_s}{\pi})E_1 - \frac{-a+b}{M^4} \bigg) \bigg] \times \bigg[ 2I^{(2)}\frac{d_{a_1}}{M^6} + 2I^{(1)}\frac{d_{a_1}}{M^4} \bigg]^{-1} , \label{kappa-sumrule}
\end{eqnarray}
\end{widetext}
where
\begin{eqnarray}
I^{(n)} & = &\int_{4 m_\pi^2}^{s_0} \rho^{\rm pole} (s) s^n ds, \nonumber
\\
E_{n} &=& 1 - e^{-s_{0}/M^{2}}\sum_{k=0}^{n}\frac{1}{k!}\bigg(\frac{s_{0}}{M^{2}}\bigg)^{k}  
\end{eqnarray}
and
$a=8\pi m_q \langle \bar{u}u \rangle$, $
b=\frac{\pi^3}{3} \langle \frac{\alpha_s}{\pi} G^2 \rangle$, $d_\rho=\frac{448 \pi^2 \alpha_s}{81 } \langle \bar{u}u \rangle^2$ and $d_{a_1}=\frac{704 \pi^2 \alpha_s}{81 } \langle \bar{u}u \rangle^2 $, $\als = 0.36$, $\langle \bar{u}u \rangle = (-0.23\; \gev)^{3}$ and $\langle \frac{\als}{\pi} G^{2}  \rangle =  (0.35 \; \gev)^{4}$.  It should be noted that although we use the vacuum saturation for the purpose of normalization, we are calculating the total value of the four quark condensates without any approximation through the $\kappa$ factors.  The  need of an effective $\kappa$ factor that is larger than 1 to correctly obtain the  $\rho$ meson mass in QCD sum rules was noticed previously in Ref. \cite{Klingl:1997kf}.

We substitute the physical meson masses and analyze the sum rule within the acceptable Borel mass range.   The lower boundary of the Borel window, $M_{\mathrm{min}}$, is determined from the condition of sufficient convergence of the OPE in Eq. ~(\ref{sum_rules}),
namely
\begin{eqnarray}
\frac{\widehat{\Pi}^{\mathrm{OPE}}_{\mathrm{cond\,terms}}(M^2)}{ \widehat{\Pi}^{\mathrm{OPE}}_{\mathrm{pert\,terms}}(M^2) } < 0.15.
\label{Borel_window_1}
\end{eqnarray}
Here, $\widehat{\Pi}^{\mathrm{OPE}}_{\mathrm{cond\,terms}}(M^2)$ is the sum of the condensate terms considered, while the denominator is the perturbative term only.  For the upper boundary, $M_{\mathrm{max}}$,
we employ the condition that the pole contribution in the integral of Eq.\,(\ref{sum_rules}) should be sufficiently large in the sum rule.
\begin{eqnarray}
\frac{  \int_{0}^{s_0} ds e^{-s/M^2} \rho(s)  }{ \int_{0}^{\infty} ds e^{-s/M^2} \rho^{OPE}(s) }
 > 0.3,
\label{Borel_window_2}
\end{eqnarray}
where the denominator is the total OPE contribution.  
The threshold is determined to obtain the most stable Borel curve.  

\begin{figure}[h]
\centerline{
\includegraphics[width=9 cm]{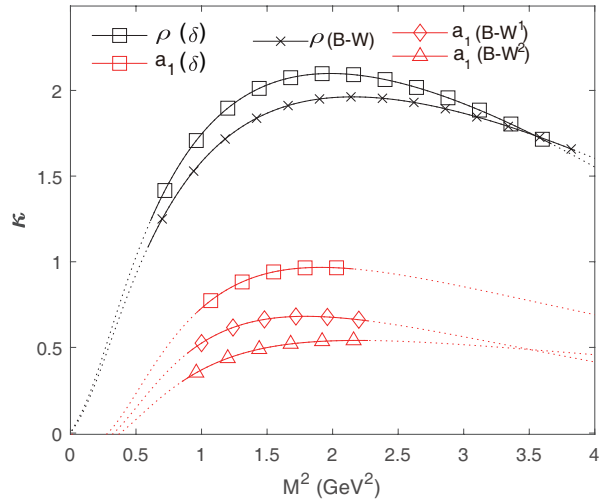}}
\caption{(Color online) Values of the four quark condensate in the $\rho$ and $a_1$ meson channel normalized to the vacuum saturation value.}
\label{kappa}
\end{figure}

\begin{figure}[h]
\centerline{
\includegraphics[width=9 cm]{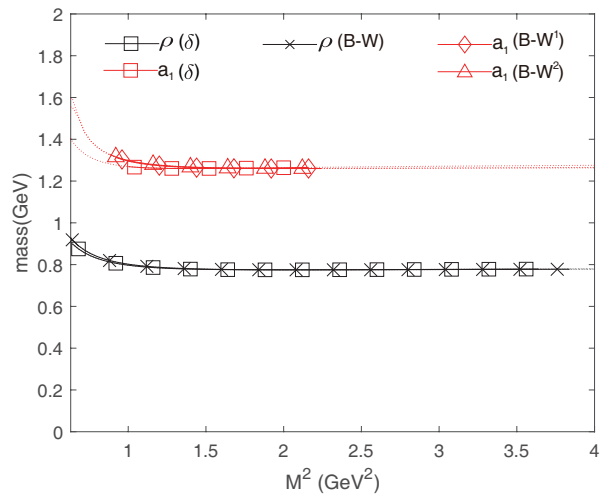}}
\caption{(Color online)  Borel curves for the mass of $\rho$ and $a_1$ meson with zero and non-zero width using the corresponding $\kappa$ values.}
\label{rho-a1-pole}
\end{figure}

Fig.~(\ref{kappa}) shows the $\kappa$ values in Eq.~(\ref{kappa-sumrule}).  The Borel windows are identified with solid lines with symbols.  The sum rule with a delta function pole, where the widths are taken to be zero (labeled $\delta$), are marked by open square symbols. The times symbols are for $\rho$ width of 150 MeV (labeled B-W) , while the diamond and triangle are for $a_1$ width of 200 (B-W$^1$) and 400 MeV (B-W$^2$), respectively.   Taking the $\kappa$ values at the extremum point, one notes that the values for the $\rho$ meson are systematically larger than those for $a_1$ meson, suggesting that the vacuum saturation is violated and that there are large contributions coming from the chiral symmetric operators. 
The different approximations are taken to asses the degree of uncertainty in our final result for the mass in the chiral symmetric vacuum. 

\begin{figure}[h]
\centerline{
\includegraphics[width=9 cm]{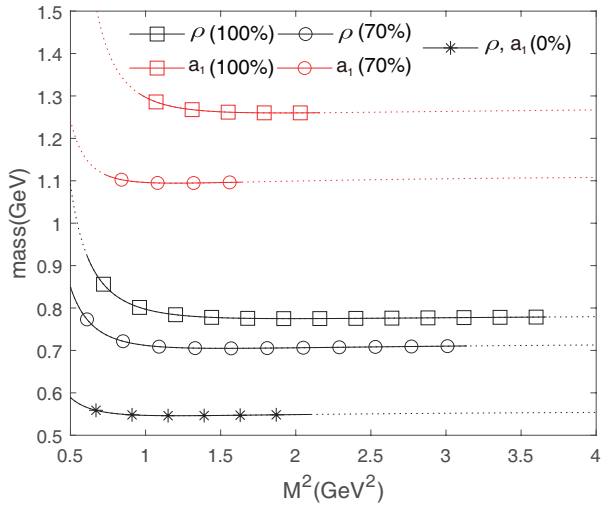}}
\caption{(Color online) Borel curves for the mass of the $\rho$ and $a_1$ meson in the vaccum (square) and when the chiral symmetry broken matrix element $B$ is reduced to 70\% (circle) and 0\% (star) of its vacuum value.}
\label{mass}
\end{figure}
 
Using these four quark condensate values, one can verify that the  corresponding Borel curve for the masses well reproduces the vacuum values. 
In the lines with solid boxes in Fig.~(\ref{rho-a1-pole}), we show the masses when the delta function approximation is taken for the poles  and their corresponding $\kappa$ values are used.  The curves with the Breit-Winger type pole contributions with the corresponding $\kappa$ values give equally good Borel curves for the masses as shown in Fig.~(\ref{rho-a1-pole}) with different symbols.  In fact, one could have used the most stable Borel curves for the masses to extract the $\kappa$ values,  with which one finds similar results.  
Table \ref{table1} summarizes the $\kappa$ values and the corresponding threshold parameters in the brackets.  Also given are the matrix elements defined in Eqs.~(\ref{rho}),(\ref{a1}) and (\ref{BS}). 

We finally show  in Fig.~(\ref{mass}), the Borel curve for the masses when we take $S$ to be the same as its vacuum value but take the chiral symmetry broken four quark condensate  $B$  to be 70\% (open circle) and  0\% (star) of its vacuum value in the delta function approximation with the threshold values of 1.12 and 0.75 GeV$^2$, respectively.  As expected, the masses of the $\rho$ and $a_1$ decrease and eventually become degenerate at around 550 MeV.
The mass at the chiral symmetry restored limit using parameter obtained with  physical widths are given in the  last column in Table. \ref{table1}.  The threshold values at the symmetric limit are 0.89 and 0.93 GeV$^2$ in the second and third row.  
When we use the $\kappa$ values obtained with non-zero widths labeled B-W$^1$ and B-W$^2$ but use the delta function ansatz for the $\rho$ pole, we find the chiral symmetric mass to be 570 and 590 MeV, respectively.  The last approximation is reasonable because when chiral symmetry is restored, the pions will become massive and restrict the phase space of the vacuum decay of both the $\rho$ and $a_1$. 
For all cases, one notes that the mass of vector meson in the chiral symmetry restored vacuum lies between 550 and 600 MeV.  

Our result, explicitly demonstrates the merging of the vector and axial vector meson mass to an  universal non vanishing value when the chiral symmetry breaking effects are restored in the vacuum.  Therefore, one can conclude that while the mass difference between chiral partners are coming from chiral symmetry breaking, the bulk part of the common mass has other non-perturbative origin.

\begin{widetext}
\begin{center}
\begin{table}[htbp]
\begin{tabular}{c c c c c c c c} \hline
	\hline
	Pole & $\kappa_{\rho}\, (s_{0})$& $\kappa_{a_{1}}\, (s_{0})$& $\mathcal{M}_{\rho}(\gev^{6})$ &$\mathcal{M}_{a_{1}}(\gev^{6})$ & $B\,(\gev^{6})$ & $S\,(\gev^{6})$ & $m_{sym}$(MeV) 	\\ 
	\hline	
	$\delta$ & 2.1(1.31)  & 0.97(2.38) & $1.72\times 10^{-3}$ & $ -1.25 \times 10^{-3}$ & $7.42 \times 10^{-4}$ & $5.65 \times 10^{-4}$ & 550 \\ 
	B-W$^{1}$ & 1.96(1.38) & 0.75(2.48) & $1.60\times 10^{-3}$ & $-0.96\times 10^{-3}$ & $6.42 \times 10^{-4}$ & $6.05 \times 10^{-4}$ & 580  
	\\ 
	B-W$^{2}$ & 1.96(1.38) & 0.54(2.5) & $1.60 \times 10^{-3}$ & $-0.69\times 10^{-3} $ & $5.75 \times 10^{-4}$ & $7.11 \times 10^{-4}$ & 600   \\
	\hline 
\end{tabular}
\caption{The four quark operators for $(\Gamma_\rho, \Gamma_{a_1})=$(0,0), (150,200)MeV  and (150,400)MeV for rows labeled ($\delta$), (B-W$^{1}$) and (B-W$^{2}$), respectively. 
The last column represents the mass in the chiral symmetric vacuum obtained with $B=0$.
Units for threshold $s_0$ is in GeV$^2$. }
\label{table1}
\end{table}
\end{center}
\end{widetext}

\section*{Acknowledgements}

This work was supported by  by Samsung Science and Technology Foundation under Project Number SSTF-BA1901-04.

\end{document}